\documentclass[12pt]{article}
\usepackage{graphicx}

\newsavebox{\sboxpubnumber}
\newsavebox{\sboxpubdate}

\newcommand{\lsim}{\mathrel{\mathop{\kern 0pt \rlap
  {\raise.2ex\hbox{$<$}}}
  \lower.9ex\hbox{\kern-.190em $\sim$}}}
\newcommand{\gsim}{\mathrel{\mathop{\kern 0pt \rlap
  {\raise.2ex\hbox{$>$}}}
  \lower.9ex\hbox{\kern-.190em $\sim$}}}
\def\Journal#1#2#3#4{{#1} {#2} (#4) {#3}}

\def\APP{\em Astrop. Phys.}

\def\NCAA{{\em Il Nuovo Cimento} A}
\def\NCC{{\em Il Nuovo Cimento} C}

\def\NIMA{{\em Nucl. Instr. \& Methods} A}
\def\NJP{\em New Journal of Physics}
\def\NPB{{\em Nucl. Phys.} B}
\def\PLB{{\em Phys. Lett.}  B}

\def\PRD{{\em Phys. Rev.} D}

\newcommand{\Title}[1]{\begin{center} {\Large #1 } \end{center}}
\newcommand{\Author}[1]{\begin{center}{ \sc #1} \end{center}}
\newcommand{\Address}[1]{\begin{center}{ \it #1} \end{center}}
\newcommand{\andauth}{\begin{center}{and} \end{center}}

\newenvironment{Abstract}{\begin{quotation}  }{\end{quotation}}
\newenvironment{Presented}{\begin{quotation} \begin{center}

             PRESENTED AT\end{center}\bigskip

      \begin{center}\begin{large}}{\end{large}\end{center}

      \end{quotation}}

\begin{document}

%%%%%%%%%%%%%%%%%%%%%%%%%%%%%%%%%%%%%%%%%%%%%%%%%%%%%%%%%%%%%%%%%%%%%%%%
%%
%% START EDITING HERE!
%%
%%%%%%%%%%%%%%%%%%%%%%%%%%%%%%%%%%%%%%%%%%%%%%%%%%%%%%%%%%%%%%%%%%%%%%%%

\begin{titlepage}
%\pubdate{\today}                    %fill in the date
%\pubnumber{XXX-XXXXX \\ YYY-YYYYYY} %preprint number(s)

\vfill

\Title{WIMP search by the DAMA experiment at Gran Sasso}

\vfill

\Author{P. Belli}

\Address{Dip.to di Fisica, Universit\`a di Roma ``Tor Vergata'' \\
and INFN, sez. Roma2, I-00133 Rome, Italy} 

\vfill
\andauth
\vfill

\Author{R. Bernabei$^a$, M. Amato$^b$, F. Cappella$^a$, R. Cerulli$^a$,
        C.J. Dai$^c$, \\ H.L. He$^c$, G. Ignesti$^b$, A. Incicchitti$^b$,
        H.H. Kuang$^c$, J.M. Ma$^c$, \\ F. Montecchia$^a$,
        F. Nozzoli$^a$, D. Prosperi$^b$
        \footnote{Neutron measurements in collaboration with: M. Angelone, P. Batistoni, M.Pillon
(ENEA, C.R. Frascati P.O. Box 65, I-00044 Frascati, Italy).
Some of the results on rare processes in collaboration with: V. Yu. Denisov, V. I. Tretyak,
O.A. Ponkratenko, Yu. G. Zdesenko (Institute for Nuclear Research, 
MSP 03680 Kiev, Ukraine).}
}

\Address{$^a$Dip.to di Fisica, Universit\`a di Roma ``Tor Vergata'' \\
        and INFN, sez. Roma2, I-00133 Rome, Italy, \\
        $^b$Dip.to di Fisica, Universit\`a di Roma ``La Sapienza'' \\
        and INFN, sez. Roma, I-00185 Rome, Italy, \\
        $^c$IHEP, Chinese Academy,  P.O. Box 918/3,
        Beijing 100039, China.}
\vfill

\begin{Abstract}
DAMA is searching for rare processes by developing and using 
several kinds of radiopure scintillators: in particular, NaI(Tl), 
liquid Xenon and CaF$_2$(Eu). The main results are here summarized 
with particular attention to the investigation of the 
WIMP annual modulation signature.
\end{Abstract}

\vfill
\begin{Presented}
    COSMO-01 \\
    Rovaniemi, Finland, \\
    August 29 -- September 4, 2001
\end{Presented}
\vfill

\end{titlepage}
\def\thefootnote{\fnsymbol{footnote}}
\setcounter{footnote}{0}

%%%%%%%%%%%%%%%%%%%%%%%%%%%%%%%%%%%%%%%%%%%%%%%%%%%%%%%%%%%%%%%%%%%%%%%%
% The document starts here
%%%%%%%%%%%%%%%%%%%%%%%%%%%%%%%%%%%%%%%%%%%%%%%%%%%%%%%%%%%%%%%%%%%%%%%%

\section{Introduction}

DAMA is devoted to the search for rare processes by developing 
and using low radioactivity scintillators. Its main aim is the search 
for relic particles (WIMPs: Weakly Interacting 
Massive Particles). 
In addition, due to the radiopurity of the used detectors and of the 
installations, 
several searches for other possible rare processes are also carried out,
such as e.g. on exotics, on $\beta\beta$
processes, on charge-non-conserving processes,
on Pauli exclusion principle violating 
processes, on nucleon instability and on solar 
axions\cite{Bel96,ncim,Bepa,Ca1,Ca2,astr,Becha,Beel,Rd99,Pl99,Ndn00,Beax,Xe136}.

The main experimental set-ups running at present are: the 
$\simeq$ 100 kg NaI(Tl) set-up, the $\simeq$ 6.5 kg liquid Xenon (LXe) 
set-up and the so-called ``R\&D'' apparatus.
Moreover, a low-background germanium detector is operative underground
for measurements and selections of sample materials.

In the following the most recent results achieved with the $\simeq$ 6.5 kg LXe 
set-up and with the ``R\&D'' apparatus will be briefly recalled, 
while a more dedicated discussion will be devoted 
to the present results on the investigation of the WIMP annual
modulation signature by the $\simeq$ 100 kg NaI(Tl) set-up.

\section{The LXe experiment}

We pointed out the interest in using liquid Xenon as target-detector 
for particle
dark matter search deep underground since ref.~\cite{Bel90}.
Several prototypes were built and related results
published~\cite{Pro}.
The final choice was to realize a pure liquid Xe scintillator 
directly collecting the emitted UV light and filled with Kr-free Xenon
gas isotopically enriched. The detailed description of this set-up and of its performances
have been described in ref. \cite{Perfo}. Kr-free Xenon enriched 
in $^{129}$Xe at 99.5\% has been used since time, while more recently the 
set-up has been modified in order to run alternatively either with this gas 
or with Kr-free Xenon enriched in $^{136}$Xe at 68.8\%.

\begin{figure}[htb]
\centering
\includegraphics[height=7.cm]{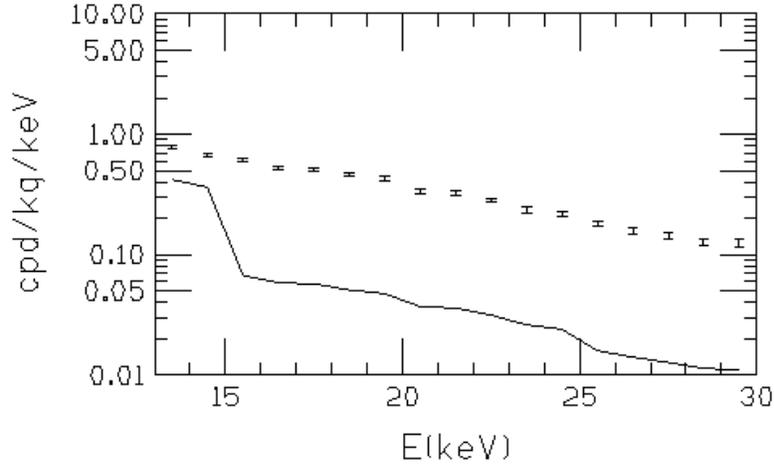}
\caption{Low energy distribution (statistics of 1763.2 $kg \times day$) 
measured with the vessel filled with $^{129}$Xe; 
the continuous line represents the upper
limits at 90\% C.L. obtained for the recoil fractions \cite{Xe98}.}
\label{fg:fig1}
\end{figure}

\begin{figure}[!b]
\centering
\vspace{-1.1cm}
\includegraphics[height=9.5cm]{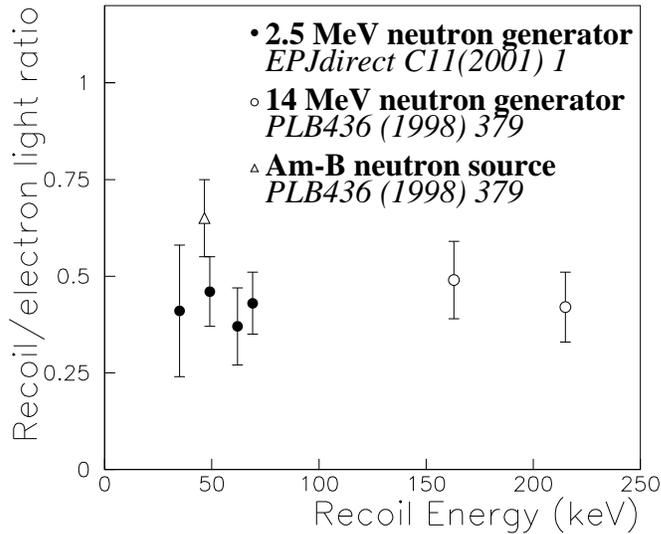}
\vspace{-1.5cm}
\caption{Measured behaviour of the recoil/electron light ratio
with recoil energy.
Note that the energy of the data point of AmB neutron source 
is an average which refers also to much lower energies than those explored 
with the 2.5 MeV neutrons.}
\label{fg:fig2}
\end{figure}

After preliminary measurements 
both on elastic and inelastic WIMP-$^{129}$Xe scattering \cite{Bei96,Bea96},
the recoil/electron light ratio and the pulse shape
discrimination capability in a similar pure LXe scintillator have been measured 
both with Am-B neutron source and with 14 MeV neutron generator \cite{Xe98}. 
After some upgrading of the set-up, 
new results on the WIMP search have been obtained~\cite{Xe98,Lxe20}.
In particular, in ref. \cite{Xe98} the pulse shape discrimination in pure LXe scintillators
has been exploited (see Fig. \ref{fg:fig1}). 
Moreover, in 2000/2001
further measurements on the recoil/electron light ratio with
2.5 MeV neutron generator have been carried out at ENEA-Frascati; 
see ref. \cite{Qf01} for details
and comparisons. Fig. \ref{fg:fig2} summarizes the measured values.
The inelastic excitation
of $^{129}$Xe by Dark Matter
particles with spin-dependent coupling has also been searched for
preliminarily in ref.~\cite{Bea96} and,
more recently, in ref.~\cite{Lxe20}.

Other rare processes have been investigated 
by filling the detector with the Kr-free Xenon gas enriched in 
$^{129}$Xe at 99.5\%.
 
In particular, as regards the electron stability, limits on the lifetime of the
electron decay in both the disappearance and the $\nu_e + \gamma$
channels were set in ref.~\cite{Bel96}. The limits on the lifetime
of the latter channel has been more
recently improved to:
$2.0(3.4) \cdot 10^{26}$ y at 90\% (68\%) C.L.~\cite{Pl99}.
Furthermore, new lifetime limits on the charge non-conserving
electron capture with excitation of $^{129}$Xe nuclear
levels have been established to be in the
range $(1-4) \cdot 10^{24}$ y at 90\% C.L.
for the different excited levels of $^{129}$Xe~\cite{Rd99}.
The most stringent restrictions on the relative strengths
of charge non-conserving processes have been derived:
$\epsilon_W^2<2.2 \times 10^{-26}$ and
$\epsilon_{\gamma}^2<1.3 \times 10^{-42}$
at 90\% C.L.

Moreover, the nucleon and di-nucleon decay into
invisible channels~\cite{Ndn00} has been investigated by
searching for the radioactive daughter nuclei, created after the
nucleon or di-nucleon disappearance in the parent nuclei.
This new approach has the advantage of a branching ratio close to 1
and -- if the parent and daughter nuclei are located
in the detector itself -- also an efficiency close to 1.
The obtained limits at 90\% C.L. are: $\tau(p \rightarrow $ invisible channel$) > 
1.9 \cdot 10^{24}$ y; $\tau(pp \rightarrow $ invisible channel$) > 5.5 \cdot 10^{23}$ y
and $\tau(nn \rightarrow $ invisible channel$) > 1.2 \cdot 10^{25}$ y.
These limits are similar or better than those previously
available; the limits for the di-nucleon decay in $\nu_\tau \bar\nu_\tau$
have been set for the first time; moreover, the limits are valid for
every possible disappearance channel~\cite{Ndn00}.

Finally as mentioned above, more recently the set-up has been modified 
to allow the use of Kr-free Xenon enriched in $^{136}$Xe at 68.8\%.
Preliminary measurements have been carried out during 6843.8 hours \cite{Xe136}
and new
experimental limits have been obtained for $\beta\beta$ decay processes
in $^{136}$Xe, improving those previously available by factors
ranging between 1.5 and 65.
In particular, for the 3 possible channels without neutrinos
the following half life limits (90\% C.L.) have been achieved:
$7.0 \cdot 10^{23}$ y for the channel $\beta\beta$ $0\nu (0^+ \rightarrow 0^+)$;
$4.2 \cdot 10^{23}$ y for the channel $\beta\beta$ $0\nu (0^+ \rightarrow 2^+)$;
$8.9 \cdot 10^{22}$ y for the channel $\beta\beta$ $0\nu M (0^+ \rightarrow 0^+)$.
For comparison, we note that the obtained experimental limits 
on the half life of the process $\beta\beta$ $0\nu (0^+ \rightarrow 0^+)$ 
is lower only than the one obtained for 
the case of $^{76}$Ge. Moreover, the limits (90\% C.L.) on the channels 
$\beta\beta$ $2\nu (0^+ \rightarrow 0^+)$
$1.1 \cdot 10^{22}$ y and $\beta\beta$ $0\nu M (0^+ \rightarrow 0^+)$ 
are at present the most stringent ones not only for the $^{136}$Xe isotopes, 
but also for every kind of nucleus investigated so far 
either by active or by passive source method.
Furthermore, upper bounds on the effective neutrino mass have been set 
considering various theoretical models for the evaluation 
of the elements of the nuclear matrix;
they vary between 1.5 eV and 2.2 eV (90\% C.L.). 
Finally, in the framework of the same
models we have also obtained 
upper limits on the effective coupling constant 
Majoron - neutrino; they range in the interval:
4.8 $\cdot 10^{-5}$ and 7.1 $\cdot 10^{-5}$ (90\% C.L.).

In conclusion, competitive results have been achieved by the 
DAMA LXe set-up by using Kr-free isotopically enriched Xenon gas.
Further upgradings to improve the detector performance are under
consideration, while the data taking is continuing with both 
enrichments.

\section{Results from the ``R\&D'' set-up}

\begin{figure}[htb]
\centering
\includegraphics[height=10.0cm]{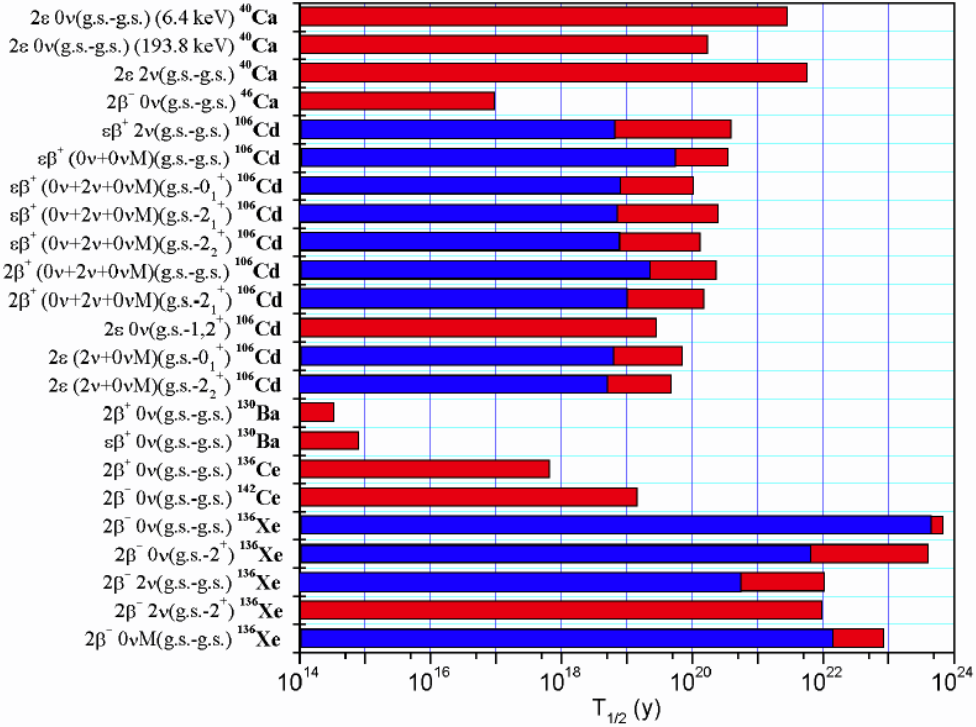}
\caption{Light gray: experimental limits on the half-life of several $\beta \beta$-decay-candidate
isotopes investigated by DAMA (90\% C.L. with exception of
the results on Barium and Cerium isotopes, which are at 68\% C.L.) \cite{ncim,Ca1,Ca2,astr}. 
Dark gray: previous best experimental limits (90\% C.L.).}
\label{fg:fig3}
\end{figure}

The ``R\&D'' set-up is used to measure the performances of
prototype detectors and to perform small scale experiments.
In particular, several experiments on the investigation of 
$\beta\beta$ decay processes in $^{136}$Ce, $^{142}$Ce, $^{40}$Ca,
$^{46}$Ca, $^{48}$Ca, $^{130}$Ba, $^{106}$Cd \cite{ncim,Ca1,Ca2,astr}
(see Fig. \ref{fg:fig3});
other measurements are in progress and/or in preparation.

\section {Results from the $\simeq$ 100 kg NaI(Tl) set-up}

The $\simeq$ 100 kg NaI(Tl) DAMA set-up allows us to explore several kinds of rare
processes regarding WIMPs, processes forbidden by Pauli exclusion 
principle, exotics, charge non
conserving processes and solar axions\cite{Bepa,Becha,Beel,Beax,Bepsd,Benc,Besimp,Beapp,Mod1,Mod2,Ext,Mod3}.
A full description of the apparatus and its performances can be found in 
ref.\cite{Beapp}.

Let us preliminarily briefly summarize recent results -- not referred to
the WIMP search -- obtained by means of this set-up.

A search for processes normally forbidden by the Pauli exclusion principle was carried 
out and an improved limit for spontaneous emission rate of protons in
$^{23}$Na and $^{127}$I has been determined. In particular, the upper limit on 
the unit time probability of non-paulian emission of protons with 
energy $E_p \ge $  10 MeV is $4.6 \times 10^{-33}$ s$^{-1}$
and the corresponding lower limit on the mean life results 
$0.7 \times 10^{25}$ y for $^{23}$Na and $0.9 \times 10^{25}$ y for
$^{127}$I \cite{Bepa}. 

The same set-up has allowed to set a new limit on the electron 
stability by considering for the first time 
the "disappearance" channel of L-shell electrons in Iodine.
With a statistics of 19511 $kg \times day$ we have obtained: 
$\tau > 4.2 \times 10^{24}$  y (68\% C.L.) for the process
$e^- \rightarrow  3\nu, majoron + \nu$ or anything invisible \cite{Beel}.

Furthermore, 
new lifetime limits have been obtained on the charge non-conserving electron 
capture with excitation of $^{127}$I and $^{23}$Na nuclear levels. 
By analysing a statistics of 34866 $kg \times day$, limits in the range: 
$(1.5-2.4) \times 10^{23}$ y (up to two orders of magnitude higher than the ones 
previously available) have been obtained; also 
stringent restrictions on the relative strengths of the charge 
non-conserving (CNC) processes have been set \cite{Becha}.

Competing results have  also been achieved by searching for neutral SIMPs:
neutral particles with masses between few GeV
and the GUT scale and cross sections on protons  up to  10$^{-22}$ cm$^2$, 
embedded  in the galactic  halo ($\beta \simeq 10^{-3}$). Although  their large  
cross section, the neutral SIMPs interaction rate should be relatively low because of their
low density in the galactic halo.
Two "planes" of NaI(Tl)  in the $\simeq$ 100 kg NaI(Tl) DAMA set-up have been used 
for this search, excluding  neutral SIMPs with masses up to 
$4 \times  10^{16}$ GeV \cite{Besimp}. With the same set-up also a new limit
 has been set on the neutral nuclearites (a possible  new  form  
of  matter containing  roughly  equal number of  up,
 down and strange quarks.). The model independent  
upper limit at 90\% C.L. on the neutral  nuclearites  flux is:
$\Phi < 1.9 \times 10^{-11}$ s$^{-1}$cm$^{-2}$sr$^{-1}$ \cite{Besimp}. 

\begin{figure}[htb]
\centering
\vspace{-1.0cm}
\includegraphics[height=8.0cm]{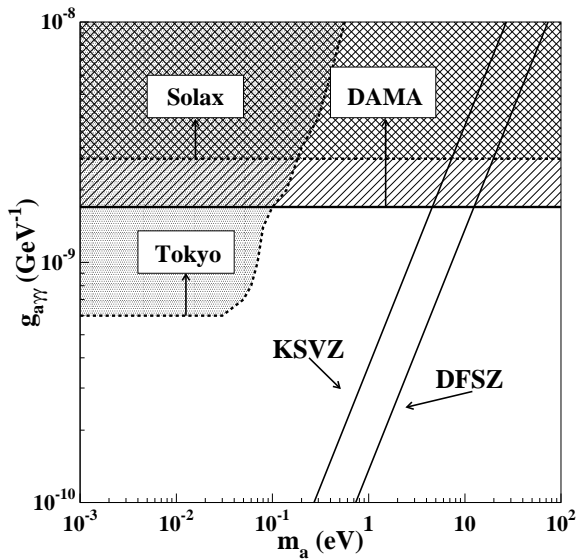}
\vspace{-0.4cm}
\caption{Exclusion plot for the axion coupling constant,
$g_{a\gamma\gamma}$, versus the axion mass, $m_a$.
The limit achieved by this experiment is shown together with 
theoretical expectations and previous direct searches 
for solar axions\cite{Beax}.}
\label{fg:axion}
\end{figure}

Recently this set-up has also allowed to set a new limit on solar axions by exploiting the 
Primakoff effect in NaI crystals. Analysing a statistics of 53437 $kg \times day$
the limit $g_{a\gamma\gamma} < 1.7 \times 10^{-9}$ GeV$^{-1}$ 
(90\% C.L.) \cite{Beax} has been obtained (see Fig.~\ref{fg:axion}); 
the region with axion mass greater than 0.03 eV (not accessible by other direct methods)
has been  explored and KSVZ axions with mass greater than 4.6 eV have been 
excluded at 90 \% C.L. (see Fig. \ref{fg:axion}).

As regards the WIMP search, the recoil/electron light ratio 
and the pulse shape discrimination
capability have been measured by neutron source and competitive results 
have been achieved on the WIMP-nucleus elastic scattering 
by exploiting the pulse shape discrimination technique 
in ref. \cite{Bepsd}. Moreover,
studies on the possible
diurnal variation of the 
low energy rate in the data collected by the $\simeq$ 100 kg NaI(Tl) have been carried out. 
We recall that a diurnal variation of the low energy rate in WIMP direct searches 
can be expected because of the Earth's daily rotation. 
In fact, during the sidereal day the Earth shields a given detector
with a variable thickness, eclipsing the ``wind'' of Dark Matter particles
but only in case of high cross section candidates (to which small 
halo fraction would correspond). By analyzing a statistics of 14962 $kg \times day$ 
no evidence of diurnal  rate variation  with sidereal time  has been observed \cite{Benc}; 
this result supports that the effect pointed out 
by the studies on the WIMP annual  modulation  signature  (see later)
would account for a halo fraction $\gsim 10^{-3}$.

Let us note that the highly radiopure $\simeq$ 100 kg 
NaI(Tl) set-up has been originally developed to investigate the
WIMP annual  modulation signature, thus the remaining part of this paper 
will be devoted to summarize the present results on this topics.

\section{Investigation of the WIMP annual modulation signature}

The WIMPs are embedded in the galactic halo; thus
our solar system, which is moving with respect to the galactic system, 
is continuously hit by a WIMP ``wind''. The 
quantitative study of such ``wind'' allows both to obtain information on the 
Universe evolution and to investigate Physics beyond the Standard Model.
The WIMPs are mainly searched for by elastic scattering on target nuclei,
which constitute a scintillation detector. In particular, the $\simeq$ 100 kg
NaI(Tl) set-up \cite{Beapp} has been realized to investigate the so-called
WIMP ``annual modulation signature''.
In fact, since the Earth rotates around the Sun, 
which is moving with respect to the galactic system,
it would be crossed by a larger WIMP flux in June (when its 
rotational velocity is summed to 
the one of the solar system with respect to the Galaxy) 
and by a smaller one in December (when the two velocities are subtracted). 
The fractional difference between the maximum and the minimum of the rate is 
expected to be of order of $\simeq$ 7\%. 
The $\simeq$ 100 kg highly radiopure NaI(Tl) DAMA set-up \cite{Beapp} 
can effectively exploit such a signature because of its
well known technology, of its high intrinsic radiopurity, of its mass, 
of the deep underground experimental site 
and of its suitable control of the operational parameters. 
The annual modulation signature is very distinctive as we have 
already pointed out \cite{Beapp,Mod1,Mod2,Ext,Mod3,Sist,Sisd,Inel}.
In fact, a WIMP-induced seasonal effect must simultaneously satisfy 
all the following requirements: the rate must contain a component 
modulated according to a cosine function (1) with one year period (2) 
and a phase that peaks around $\simeq$ 2$^{nd}$ June (3); 
this modulation must be found 
in a well-defined low energy range, where WIMP induced recoils 
can be present  (4); it must apply to those events in 
which just one detector of many actually "fires", since
the WIMP multi-scattering probability is negligible (5); the modulation
amplitude in the region of maximal sensitivity must be $\lsim$7$\%$ (6). 
Only systematic effects able 
to fulfil these 6 requirements could fake this signature;
therefore for some other effect 
to mimic such a signal is highly unlikely. 

Here the results obtained by analysing the data of four annual
cycles investigating the annual modulation signature by means of 
the $\simeq$ 100 kg NaI(Tl) set-up are summarized (see Table \ref{noi}).
\begin{table}[!h] 
\caption{Released data sets \cite{Bepsd,Mod1,Mod2,Ext,Mod3,Sisd,Inel}.}
\label{noi}
\begin{center}
\begin{tabular}{cc}
period  & statistics ($kg \times day$) \\
\hline
DAMA/NAI-1        &  4549 \\
DAMA/NaI-2        & 14962 \\
DAMA/NaI-3        & 22455 \\
DAMA/NaI-4        & 16020 \\
Total statistics  & 57986 \\
\hline
+ DAMA/NaI-0 &  upper limit on recoils by PSD \\
\hline
\end{tabular}
\end{center}
\end{table}       
Cumulative analyses of 
a total statistics of 57986 $kg \times day$ 
have been carried out \cite{Mod3,Sist,Sisd} properly including
the physical constraint which arises from the 
measured upper limit on the recoil rate \cite{Bepsd,Sist}. 
Both model independent and model dependent analyses have been
performed.
We remark that 
various uncertainties exist in every model dependent 
calculations concerning WIMP
direct detection search
(the same is for exclusion plot 
as well as for allowed regions in the WIMP-proton cross section versus
WIMP mass plane). 
They are present on some general features such as e.g. 
the real behaviour of the WIMP velocity distribution 
(which we have considered so far isothermal and Maxwellian), 
on the adopted particle model and on the values taken for all the parameters needed 
in the calculations.

\section{Annual modulation signature and related studies}

\vskip 0.2cm 

\subsection{Model independent analysis}

\begin{figure}[htb]
\centering
\includegraphics[height=7.cm]{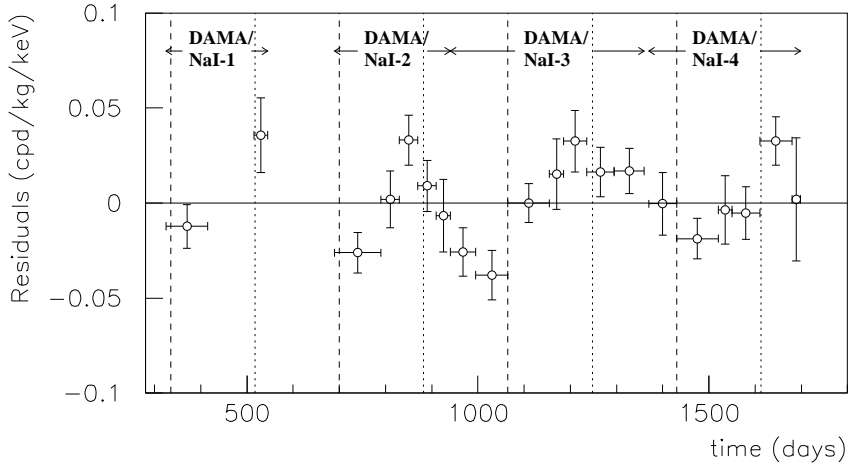}
\caption{Model independent residual rate for single hit events, in the 
2--6 keV cumulative energy interval, as a function of the time elapsed since
January 1-st of the first year of data taking. The expected
behaviour of a WIMP signal is
a cosine function with minimum roughly at the dashed
vertical lines and with maximum roughly at the dotted ones.}
\label{fg:fig5}
\end{figure}

A model independent analysis of the data of the four annual cycles 
offers an immediate evidence of the presence of 
an annual modulation of the rate of the single hit events
in the lowest energy interval (2 -- 6 keV) as shown in Fig. \ref{fg:fig5}. 
There each data point has been obtained 
from the raw rate measured in the corresponding time interval, 
after subtracting the constant part.

The $\chi^2$ test on the data of Fig. \ref{fg:fig5}
disfavors the hypothesis of unmodulated behaviour
(probability: $4 \cdot 10^{-4}$), while 
fitting these residuals with the function
$A \cdot$ cos$\omega (t-t_0)$, one gets:
i) for the period T = $\frac{2\pi}{\omega}$ = (1.00 $\pm$ 0.01) year
when $t_0$ is fixed at the 152.5$^{th}$ day of the year (corresponding to
$\simeq$ 2 June); ii) 
for the phase t$_0$ = (144 $\pm$ 13) days,
when T is fixed at 1 year. In the two cases $A$ is:
(0.022 $\pm$ 0.005) cpd/kg/keV and 
(0.023 $\pm$ 0.005) cpd/kg/keV, respectively.

We have extensively discussed the results of the investigations
of the  known sources of possible systematics when releasing 
the data of each annual cycle; moreover, a dedicated 
paper \cite{Sist} has been released on possible systematics,
where in particular
the data of the DAMA/NaI-3 and DAMA/NaI-4 running periods 
have been considered in quantitative evaluations. No known 
systematic effect or side reaction able to mimic a WIMP induced 
effect has been found as discussed in details in ref. \cite{Sist}.

In conclusion, a WIMP contribution 
to the measured rate is candidate by the result of the model independent
approach independently on the nature and coupling 
with ordinary matter of the possible WIMP particle. 

\vskip 0.2cm 

\subsection{Model dependent analyses}

To investigate the nature and coupling
with ordinary matter of a possible candidate, a suitable energy and time 
correlation analysis is necessary as well as a complete model 
framework. We remark that a model framework is identified 
not only by general 
astrophysical, nuclear and particle physics assumptions, but also by the set 
of values used for all the parameters needed in the model itself and
in related quantities 
(for example WIMP local velocity, $v_0$, form factor parameters, etc.).

At present the lightest supersymmetric
particle named neutralino is considered the best candidate for WIMP.
In supersymmetric theories 
both the squark and the Higgs bosons exchanges give contribution to
the coherent (SI) part of the neutralino cross section, while
the squark and the $Z^{0}$ exchanges give contribution
to the spin dependent (SD) one.
Note, in particular, that the 
results of the data analyses \cite{Mod3,Sisd,Inel} 
summarized here and in the 
following hold for the neutralino, but are not restricted only to
this candidate.

The differential energy distribution of the recoil nuclei 
in WIMP-nucleus elastic scattering
can be calculated \cite{Bepsd,Botdm}
by means of the differential cross section of the WIMP-nucleus elastic 
processes
\begin{eqnarray}
\frac{d\sigma}{dE_R}(v,E_R) = \left( \frac{d \sigma}{dE_{R}} \right)_{SI}+
\left( \frac{d \sigma}{dE_{R}} \right)_{SD}
\label{eq:prosezdiff1}
\end{eqnarray}   
where $v$ is the WIMP velocity in the laboratory frame and
$E_R$ is the recoil energy.

In the following, the results obtained by analysing the data in some of the
possible model frameworks are summarized.

\vskip 0.2cm
{\bf I. WIMPs with dominant SI interaction in a given model framework}
\vskip 0.2cm 

\normalsize

Often the spin-independent interaction 
with ordinary matter is assumed to be dominant since e.g.
most of the used target-nuclei are practically
not sensitive to SD interactions as on the contrary 
$^{23}$Na and $^{127}$I are.
Therefore, first model dependent
analyses of the data have been performed by considering a candidate 
in this scenario, that is neglecting the term 
$\left( \frac{d \sigma}{dE_{R}} \right)_{SD}$
in eq. (1).

\begin{figure}[htb]
\centering
\includegraphics[angle=90,height=6.cm]{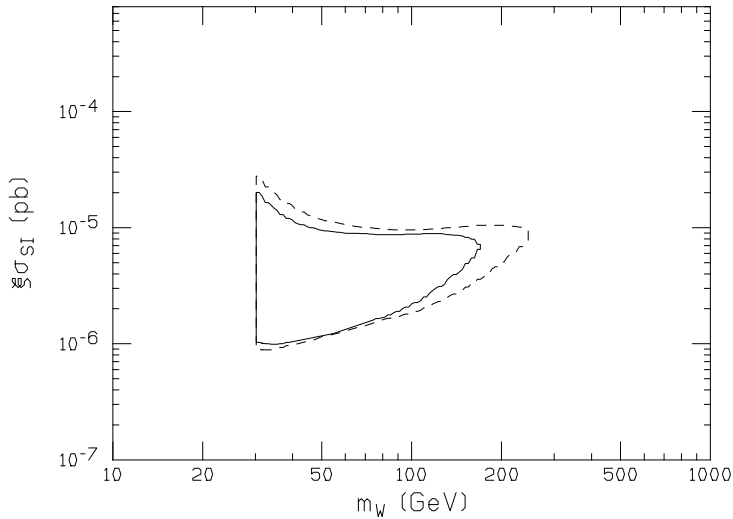}
\caption{Regions allowed at 3$\sigma$ C.L. on the plane 
$\xi\sigma_{SI}$ versus m$_W$
for a WIMP with dominant SI interaction and mass above 30 GeV in the 
model framework considered in ref. \cite{Mod3}:
i) when $v_0$ uncertainty (170 km/s $\le v_0 \le 270$ km/s;
continuous contour) has been included; ii) when 
also a possible bulk halo rotation as in ref. \cite{Ext} (dashed contour) is 
considered. As widely known, 
the inclusion of present uncertainties on some other astrophysical, 
nuclear and particle 
physics parameters would enlarge these regions (varying consequently the
position of the minimum for the $y$ log-likelihood function).} 
\label{fg:fig6}
\end{figure}

A full energy and time correlation 
analysis -- properly accounting for the physical constraint arising 
from the measured upper limit on recoils \cite{Bepsd,Sist}
-- has been carried out in the framework of a given 
model for spin-independent coupled candidates with mass above 30 GeV.
A standard 
maximum likelihood method has been used\footnote{Substantially the 
same results are obtained with other analysis approaches 
such as e.g. the Feldman and Cousins one.}.
Following the usual procedure we have built the $y$ log-likelihood function,
which depends on the experimental data and on the 
theoretical expectations;
then, $y$ is minimized and parameters' regions allowed 
at given confidence level are derived. 
Note that different model frameworks (see above)
vary the expectations and, therefore, the cross section and mass values
corresponding to the $y$ minimum, that is also the allowed region
at given C.L.. In particular, 
the inclusion of the uncertainties associated to the models and to 
every parameter in the models themselves
as well as other possible scenarios 
enlarges the allowed region
as discussed e.g. in ref. \cite{Ext}
for the particular case of the astrophysical velocities. 
Also in the case considered here the minimization procedure 
has been repeated by varying the WIMP local velocity, $v_0$, from 170 km/s to 270 km/s 
to account for its present uncertainty.
For example, in the model framework considered in ref. \cite{Mod3}, the values 
$m_W = (72^{+18}_{-15})$ GeV and
$\xi \sigma_{SI} = (5.7 \pm 1.1) \cdot 10^{-6}$ pb
correspond to the position of $y$ minimum when $v_0$ = 170 km/s,
while $m_W = (43^{+12}_{-9})$ GeV and
$\xi \sigma_{SI} = (5.4 \pm 1.0) \cdot 10^{-6}$ pb
when $v_0$ = 220 km/s. Here, $\xi$ is the WIMP local density
in 0.3 GeV cm$^{-3}$ unit,
$\sigma_{SI}$ is the point-like SI WIMP-nucleon 
generalized cross section and $m_W$ is the WIMP mass.
Fig. \ref{fg:fig6} shows the regions allowed at 3$\sigma$ 
C.L. in such a model framework,
when the uncertainty on $v_0$ is taken into account (solid contour) and
when possible bulk halo rotation is considered (dashed contour).  
For simplicity, no other uncertainties on the used parameters have been 
considered there
(some of them have been included in the approach summarized in the next 
subsection \cite{Sisd}), which obviously further enlarge the allowed region
with the respect both to $\xi \sigma_{SI}$ and $m_W$.

A quantitative comparison between the results of the model independent 
analysis and of this  model dependent analysis has been discussed 
in ref. \cite{Mod3}.

In conclusion, the observed effect investigated in terms of
a WIMP candidate with dominant SI interaction and mass above 30 GeV
in the model framework considered in ref. \cite{Mod3},
supports allowed WIMP masses up to 130 GeV (1 $\sigma$ C.L.) and even 
up to 180 GeV (1 $\sigma$ C.L.) if possible dark halo rotation is included,
while lower $\xi\sigma_{SI}$ would be implied by the inclusion of known 
uncertainties on parameters (for example, on the form factor parameters) 
and model features.

Theoretical implications of these results in terms 
of a neutralino with dominant SI interaction and mass above 30 GeV
have been discussed in ref. \cite{Botdm,Botfg,Arn},
while the case for an heavy neutrino of the fourth family 
has been considered in ref. \cite{Far}.

\vskip 0.2cm
{\bf II. WIMPs with mixed coupling in given model framework} 
\vskip 0.2cm
\normalsize   

Since the $^{23}$Na and $^{127}$I nuclei are sensitive 
to both SI and SD couplings -- on the contrary e.g. of 
$^{nat}$Ge and $^{nat}$Si which are sensitive mainly 
to WIMPs with SI coupling (only 7.8 \% is non-zero spin isotope in $^{nat}$Ge
and only 4.7\% of $^{29}$Si in $^{nat}$Si) --
the analysis of the data has been extended 
considering the more general case of eq. (1) \cite{Sisd}.
This implies a WIMP having not only a spin-independent, 
but also a spin-dependent coupling different from zero, as it is also 
possible e.g. for the neutralino (see above).

Following the usual procedure we have built the $y$ log-likelihood function,
which depends on the experimental data and on the 
theoretical expectations in the given model framework.
Then, $y$ has been minimized -- properly accounting also for the physical 
constraint set by the measured upper limit on recoils \cite{Bepsd} --
and parameters' regions allowed 
at given confidence level have been obtained.
In particular, the calculation has been performed
by minimizing the $y$ function with respect to
the $\xi \sigma_{SI}$, $\xi \sigma_{SD}$ and $m_W$
parameters for each given $\theta$ value.
Here,  $\sigma_{SD}$  is the point-like SD WIMP
cross section on nucleon and $tg\theta$ is the ratio
between the effective SD coupling constants on neutrons, $a_n$,
and on proton, $a_p$; therefore, $\theta$ can assume values between
0 and $\pi$ depending on the SD coupling.
In the present framework the uncertainties on $v_0$ have been included;
moreover, the uncertainties on the nuclear 
radius and the nuclear surface thickness parameter in the SI form factor, on 
the $b$ parameter in the used SD form factor (see later)
and on the measured quenching factors \cite{Bepsd} of these detectors
have also been considered \cite{Sisd}. 

\begin{figure}[htb]
\centering
\includegraphics[height=10.cm]{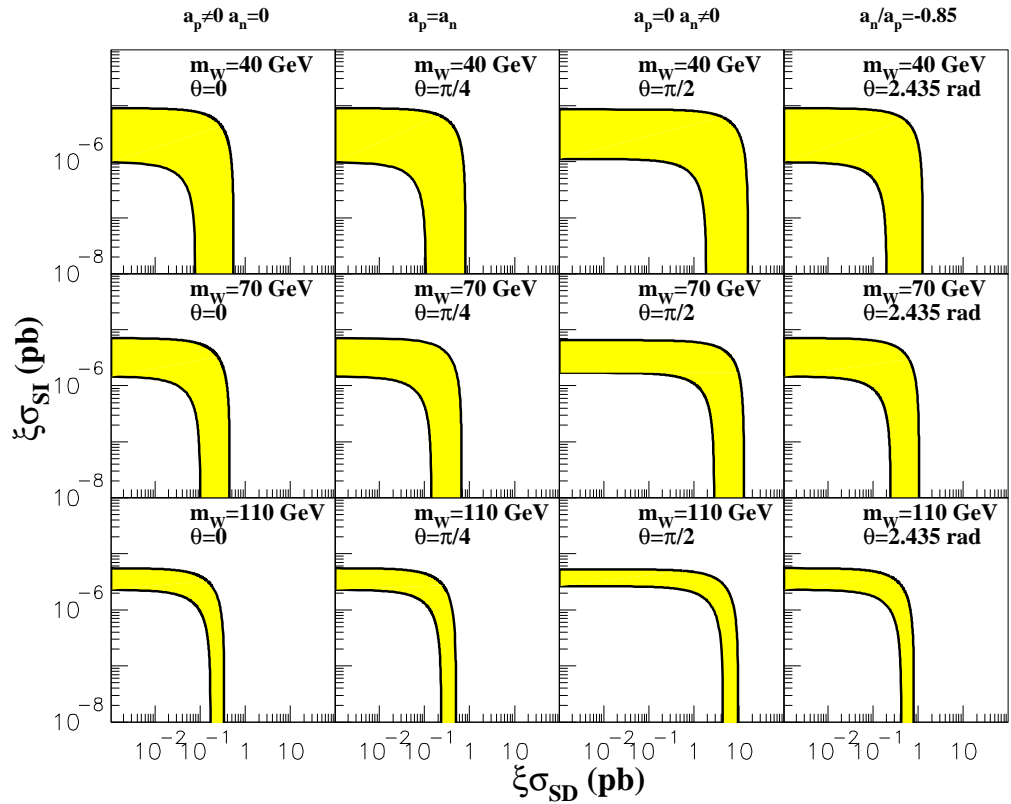}
\caption{Slices for some 
$m_W$ of the region allowed at 3 $\sigma$ C.L. 
in the ($\xi \sigma_{SI}$, $\xi \sigma_{SD}$, $m_W$) space for some $\theta$
value in the model framework considered in ref.~\cite{Sisd}. 
Only four particular couplings are reported here for simplicity:
i) $\theta$ = 0;
ii) $\theta$ = $\pi/4$ 
iii) $\theta$ = $\pi/2$; iv) 
$\theta$ = 2.435 rad.}
\label{fg:fig7}
\end{figure}

Fig. \ref{fg:fig7} shows slices for some 
$m_W$ of the region allowed at 3 $\sigma$ C.L. 
in the ($\xi \sigma_{SI}$, $\xi \sigma_{SD}$, $m_W$) space 
for some $\theta$ value.
Only the case of 
four particular couplings are shown here for simplicity:
i)  $\theta$ = 0 ($a_n$ =0 and $a_p \ne$ 0 or $|a_p| >> |a_n|$); 
ii) $\theta = \pi/4$ ($a_p = a_n$);
iii)  $\theta$ = $\pi/2$ ($a_n \ne$ 0 and $a_p$ = 0 or  $|a_n| >> |a_p|$); 
iv) $\theta$ = 2.435 rad ($ \frac {a_n} {a_p}$ = -0.85, pure Z$^0$ coupling).
The case $a_p = - a_n$ is nearly similar to the case iv). 

Comments about other discussions of  purely SD component
as well as about the
comparisons with other direct and indirect experiments 
can be found in ref. \cite{Sisd}.

As already pointed out, when the SD contribution goes to zero 
(y axis in Fig. \ref{fg:fig7}), an interval not 
compatible with zero is obtained for $\xi\sigma_{SI}$.
Similarly, when the SI contribution goes to zero
(x axis in Fig. \ref{fg:fig7}), finite values for the SD cross section are obtained. 
Large regions are allowed for mixed configurations
also for $\xi\sigma_{SI} \lsim 10^{-5}$ pb and
$\xi\sigma_{SD} \lsim 1$ pb; only in the particular case of 
$\theta = \frac {\pi} {2}$ (that is $a_p = 0$ and $a_n \ne 0$)
$\xi\sigma_{SD}$ can increase up to $\simeq$ 10 pb, since 
the $^{23}$Na and $^{127}$I nuclei have the proton as odd
nucleon. Moreover, in ref. \cite{Sisd} we have also pointed out that: 
i) finite values can be allowed for $\xi\sigma_{SD}$ 
even when $\xi\sigma_{SI} \simeq 3 \cdot 10^{-6}$ pb as in the region allowed 
in the pure SI scenario considered in the previous subsection;
ii) regions not compatible with zero 
in the $\xi\sigma_{SD}$ versus m$_W$ plane are allowed 
even when $\xi\sigma_{SI}$ values much lower than those allowed in 
the dominant SI scenario previously summarized are considered; iii) 
minima of the $y$ function with both $\xi\sigma_{SI}$ and
$\xi\sigma_{SD}$ different from zero are present for some $m_W$ and
$\theta$ pairs; the related confidence level ranges between 
$\simeq$ 3 $\sigma$ and $\simeq$ 4 $\sigma$ \cite{Sisd}.

In conclusion, this analysis has shown that the DAMA data 
of the four annual cycles, analysed in terms
of WIMP annual modulation signature,
can also be compatible with a mixed scenario 
where both $\xi\sigma_{SI}$ and $\xi\sigma_{SD}$ are
different from zero. The pure SD and pure SI cases in the 
model framework considered here 
are implicitly given in Fig.  \ref{fg:fig7} for the quoted $m_W$ and $\theta$ values.

Further investigations are in progress on these
model dependent analyses to account for other known parameters uncertainties
and for possible different model assumptions.
As an example we recall 
that for the SD form factor 
an universal formulation is not possible since 
the internal degrees of the WIMP particle model (e.g. supersymmetry
in case of neutralino) cannot be completely separated
from the nuclear ones. In the calculations presented here we have adopted
the SD form factors of ref. \cite{Ress97} estimated by considering
the Nijmengen nucleon-nucleon potential. Other formulations are 
possible for SD form factors and can be considered with evident implications
on the obtained allowed regions.  

\vskip 0.2cm
{\bf III. Inelastic Dark matter}
\vskip 0.2cm
\normalsize   

It has been recently suggested  \cite{Wei01} the possibility that the observed 
annual modulation of the low energy rate could be induced by possible 
inelastic Dark Matter: relic particles that cannot scatter elastically 
off of nuclei. As discussed in ref. \cite{Wei01}, the inelastic Dark Matter 
could arise from a massive complex scalar split into two approximately 
degenerate real scalars or from a Dirac fermion split into two 
approximately degenerate Majorana fermions, namely $\chi_+$ and $\chi_-$,
with a $\delta$ mass splitting. In particular, a specific
model featuring a real component of the sneutrino,
in which the mass splitting naturally arises, has been given in ref.
\cite{Wei01}.
It has been shown that for the $\chi_-$ inelastic scattering 
on target nuclei a kinematical constraint exists which favours
heavy nuclei (such as $^{127}$I) with respect to 
lighter ones (such as e.g. $^{nat}$Ge) as target-detectors media.
In fact, $\chi_{-}$ can only inelastically scatter
by transitioning to $\chi_{+}$ (slightly heavier state than $\chi_{-}$)
and this process can occur
only if the $\chi_{-}$ velocity, $v$, is larger than:
\begin{equation}
v_{thr} = \sqrt{\frac{2\delta}{m_{WN}}},
\label{eq:constraint}
\end{equation}
where $m_{WN}$ is the WIMP-nucleus reduced mass and hereafter $c=1$.
This kinematical constraint becomes increasingly severe 
as the nucleus mass, $m_N$, is decreased \cite{Wei01}. 
Moreover, this model scenario gives rise -- with respect to the
case of WIMP elastically scattering -- to an enhanced 
modulated component, $S_m$, with respect to the unmodulated one, $S_0$,
and to largely different behaviours with energy for
both $S_0$ and $S_m$ (both show a higher mean value) \cite{Wei01}.

For the sake of completeness, we remind that -- as 
stressed in ref. \cite{Wei01} -- this scenario is suitable to reconcile
in every case  
the DAMA result \cite{Mod1,Mod2,Ext,Mod3,Sist,Sisd,Inel} 
with the CDMS-I claim \cite{CDMS}; however, it is worth to note that
-- as discussed e.g. in sect. III of ref. \cite{texas} -- in reality the 
claim for contradiction made by CDMS-I was largely unjustified 
both for experimental and theoretical reasons.  

Anyhow, the proposed inelastic Dark Matter scenario \cite{Wei01}
offers a further possible model framework, which has also the merit to 
recover the sneutrino as a suitable WIMP candidate.

Therefore, a dedicated energy and time
correlation analysis of the DAMA experimental data has been carried out \cite{Inel}
by considering the data collected during four annual cycles 
(statistics of 57986 $kg \times day$)
\cite{Mod1,Mod2,Ext,Mod3,Sist,Sisd,Inel}.
Here aspects other than the interaction type
have been handled according to ref. \cite{Sisd}, fixing in this 
way a given model framework.

\begin{figure}[htb]
\centering
\includegraphics[height=9.cm]{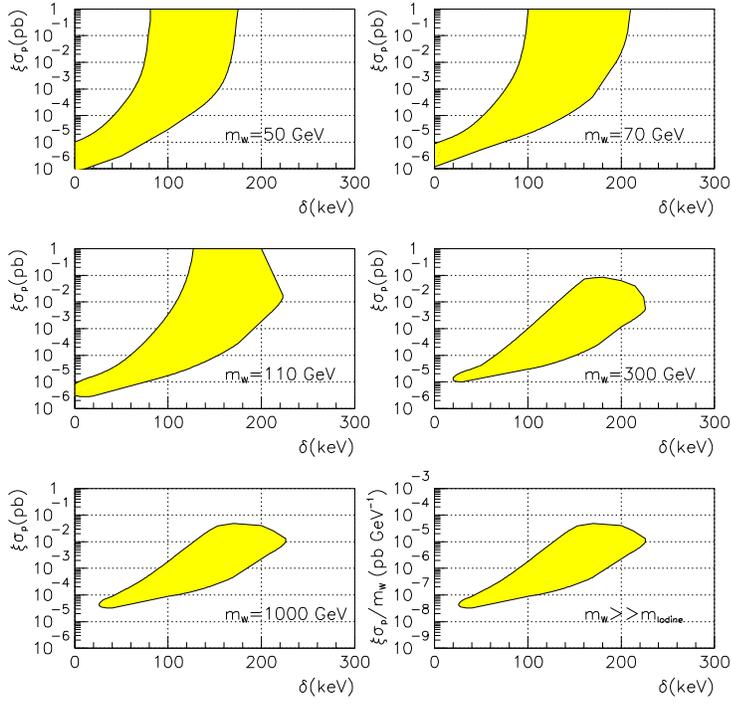}
\caption{Slices at fixed WIMP masses of the volume 
allowed at 3 $\sigma$ C.L. in the space ($\xi \sigma_p$, $\delta$, $m_W$); the 
uncertainties on some of the used parameters have been included \cite{Inel}. See text.} 
\label{fg:fig8}
\end{figure}

In this inelastic Dark Matter scenario an allowed volume 
in the space ($\xi \sigma_p$,$m_W$,$\delta$) 
is obtained \cite{Inel}. For simplicity, Fig. \ref{fg:fig8} shows slices of such an allowed 
volume at some given WIMP masses (3 $\sigma$ C.L.). 
It can be noted that when $m_W \gg m_N$, 
the expected differential energy spectrum is trivially dependent on $m_W$ 
and in particular it is proportional to the ratio between $\xi \sigma_p$ 
and $m_W$; this particular case is summarized in the last
plot of Fig. \ref{fg:fig8}.
The allowed regions reported there have been obtained by the superposition
of those obtained when varying the values of the previously 
mentioned parameters according to their uncertainties.
This also gives as a consequence that the cross section value 
at given $\delta$ can span there over several orders of magnitude.
The upper border of each region is reached when 
$v_{thr}$ approximates the maximum WIMP velocity in the Earth frame
for each considered model framework (in particular, for each 
$v_0$ value).

Note that each set of values (within those allowed by 
the associated uncertainties) for the previously mentioned 
parameters gives rise to a different expectation, thus to different 
best fit values. 
As an example we mention that when fixing 
the other parameters as in Ref. \cite{Sisd},
the best-fit values for a WIMP mass of 70 GeV 
are: i) $\xi \sigma_p$ = $2.5 \times
10^{-2}$ pb and $\delta = 115$ keV when $v_0 = 170$ km/s,
ii) $\xi \sigma_p$ = $6.3 \times
10^{-4}$ pb and $\delta = 122$ keV when $v_0 = 220$ km/s.

Finally, we note also here that significant enlargement of the given allowed regions 
should be expected when including complete effects of model (and related 
experimental and theoretical parameters) uncertainties.
Moreover, possible different halo models can be also considered.

\subsection{Conclusions}

The DAMA annual modulation data of four annual cycles 
\cite{Mod1,Mod2,Ext,Mod3,Sist,Sisd,Inel}
have been analysed by energy and time correlation analysis
in terms of purely SI, purely SD, mixed SI/SD, 
``preferred'' inelastic WIMP scattering model frameworks.

To effectively discriminate among the different possible scenarios
further investigations are in progress.
In particular, the data of the 5$^{th}$ and 
6$^{th}$ annual cycles are at hand, while the set-up is running to collect 
the data of a 7$^{th}$ annual cycle. Moreover, the LIBRA (Large sodium 
Iodine Bulk for RAre processes) set-up is under 
construction to increase the experimental sensitivity.

\section{Towards LIBRA}

At present our main efforts are devoted to the realization
of LIBRA (Large sodium Iodine Bulk for RAre processes in the DAMA
experiment) consisting of $\simeq$ 250 kg of NaI(Tl). New radiopure
detectors by chemical/physical purification of NaI and Tl powders as a
result of a dedicated R$\&$D with Crismatec have been realized.
This will allow to increase the sensitivity of
the experiment.

\end{document}